\documentclass[conference]{IEEEtran}
\IEEEoverridecommandlockouts
\usepackage{hyperref}
\usepackage{diagbox}
\usepackage{balance}
\usepackage{graphicx}
\usepackage{textcomp}
\usepackage{xcolor}
\usepackage{caption}
\usepackage{makecell}
\usepackage{booktabs}
\usepackage{subcaption}
\usepackage[most]{tcolorbox}
\usepackage{afterpage}
\usepackage{algpseudocode}
\usepackage{algorithm}
\usepackage{amsfonts}

\usepackage{enumitem}

\newtcolorbox{mybox}{
enhanced,
boxrule=0pt,frame hidden,
borderline west={4pt}{0pt}{green!75!black},
colback=green!10!white,
sharp corners
}

\def\BibTeX{{\rm B\kern-.05em{\sc i\kern-.025em b}\kern-.08em
    T\kern-.1667em\lower.7ex\hbox{E}\kern-.125emX}}
\begin{document}

\title{Coordinated Power Management on Heterogeneous Systems
\thanks{This work will present in ACM ICS 2026.}
}

\author{
\IEEEauthorblockN{Zhong Zheng}
\IEEEauthorblockA{\textit{Department of Computer Science} \\
\textit{University of Illinois Chicago} \\
zzheng33@uic.edu}
\and
\IEEEauthorblockN{Zhiling Lan}
\IEEEauthorblockA{\textit{University of Illinois Chicago} \\
\textit{Argonne National Laboratory} \\
zlan@uic.edu}
\and
\IEEEauthorblockN{Xingfu Wu}
\IEEEauthorblockA{\textit{Mathematics and Computer Science} \\
\textit{Argonne National Laboratory} \\
xingfu.wu@anl.gov}
\and
\IEEEauthorblockN{Valerie E. Taylor}
\IEEEauthorblockA{\textit{Mathematics and Computer Science} \\
\textit{Argonne National Laboratory} \\
vtaylor@anl.gov}
\and
\IEEEauthorblockN{Michael E. Papka}
\IEEEauthorblockA{\textit{Argonne National Laboratory} \\
\textit{University of Illinois Chicago} \\
papka@anl.gov}
}

\maketitle
\begin{abstract}
Performance prediction is essential for energy-efficient computing in heterogeneous computing systems that integrate CPUs and GPUs. However, traditional performance modeling methods often rely on exhaustive offline profiling, which becomes impractical due to the large setting space and the high cost of profiling large-scale applications. In this paper, we present OPEN, a framework consists of offline and online phases. The offline phase involves building a performance predictor and constructing an initial dense matrix. In the online phase, OPEN performs lightweight online profiling, and leverages the performance predictor with collaborative filtering to make performance prediction. We evaluate OPEN on multiple heterogeneous systems, including those equipped with A100 and A30 GPUs. Results show that OPEN achieves prediction accuracy up to 98.29\%. This demonstrates that OPEN effectively reduces profiling cost while maintaining high accuracy, making it practical for power-aware performance modeling in modern HPC environments. Overall, OPEN provides a lightweight solution for performance prediction under power constraints, enabling better runtime decisions in power-aware computing environments.
\end{abstract}


\section{Introduction}\label{Introduction}

As high-performance computing (HPC) systems continue to grow in scale and complexity, energy efficiency has become a critical design and operational concern \cite{bergman2008exascale}. A significant portion of total system power is consumed by processors and accelerators, necessitating the need for effective power management strategies. Cantalupo et al. \cite{strawman} propose PowerStack, a hierarchical framework for power and energy management in HPC systems. PowerStack defines multiple layers of management from a system perspective, including cluster-wide power budgeting and application-level power budgeting. Cluster-wide power budgeting ensures that the total power consumption of all running workloads does not exceed the system’s overall power budget. Application-level power budgeting, on the other hand, focuses on selecting power settings for individual applications to maximize energy efficiency. The application-level of control is essential for cluster-wide power management: for instance, by reducing power consumption at the application level, more applications can run concurrently within the same fixed cluster-level power budget. Therefore, it is vital for the system to quickly identify an application's performance characteristics under a wide range of power settings and select the optimal energy-efficient setting when the applications is submitted to system for execution. In this work, we focus on the application-level energy-efficiency. We define energy efficiency as minimizing the power consumption required to complete a unit of work in this paper.

The evolution of HPC systems has witnessed a significant shift toward heterogeneous CPU-GPU architectures (one or more CPUs paired with one or more GPUs). This transformation has been accompanied by the emergence of diverse GPU applications in HPC environments, ranging from traditional scientific computing workloads utilizing CUDA, OpenCL, and SYCL frameworks to modern neural network training that leverage GPU parallelism \cite{bergman2008exascale, che2009rodinia}. However, these GPU applications exhibit varying computational characteristics and resource utilization patterns. While some applications are predominantly GPU-bound due to their highly parallel nature and intensive floating-point operations, others remain CPU-bound owing to sequential processing requirements, complex control flow, or frequent CPU-GPU data transfers \cite{hong2010integrated, lee2011improving}. Given this heterogeneity in application behavior and resource sensitivity, effective power management strategies for modern HPC systems must account for the diverse characteristics of GPU applications by considering both CPU and GPU power management. 

Plenty of power management works have been done to improve energy efficiency for GPU applications. For instance, power capping and frequency scaling are widely adopted techniques to improve energy efficiency \cite{dutta2018gpu, fan2019predictable, wang2020gpgpu, guerreiro2019dvfs, wu2015gpgpu, ali2023performance, guerreiro2018gpgpu, dvfs}. Existing works can be broadly classified into three categories: offline-only \cite{offline-1, offline-2, offline-3, offline-4, offline-5, offline-6, offline-7, offline-8, offline-9, offline-10, offline-11}, hybrid \cite{hybrid-1, hybrid-2, hybrid-3, hybrid-4, hybrid-5, hybrid-6, hybrid-7, hybrid-8, hybrid-9, hybrid-10, hybrid-11, hybrid-12, sim-1, sim-2, sim-3}, and online-only \cite{online-1, online-2} approaches. 
\begin{figure*}
    \centering
    \includegraphics[width=1\linewidth]{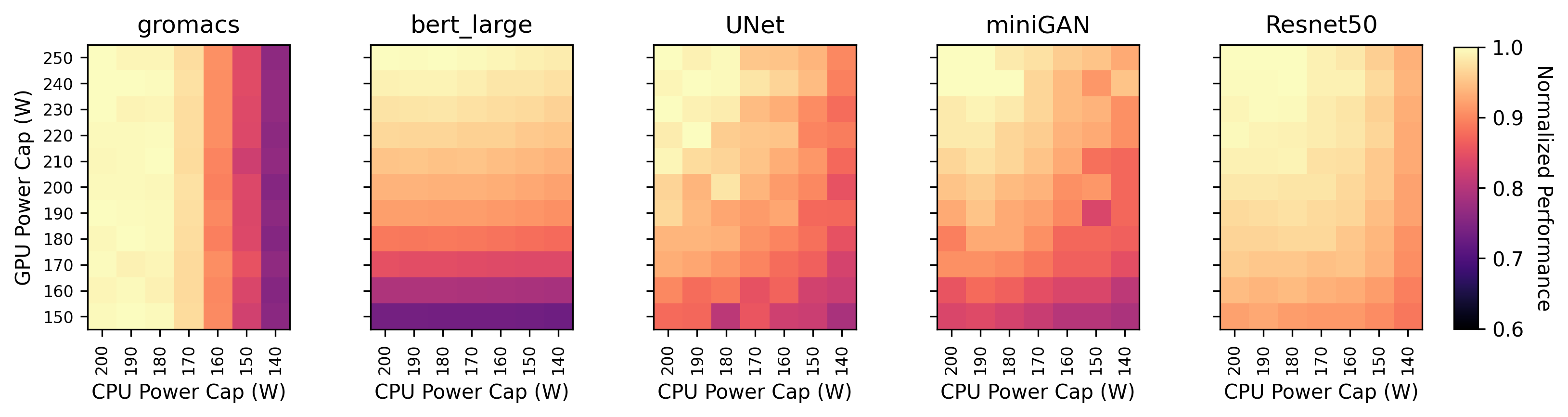}
    \caption{Normalized performance for \textit{BERT} training, \textit{UNet} training, \textit{miniGAN}, \textit{GROMACS}, and \textit{Resnet50} training on an Intel(R) Xeon(R) Platinum 8380 system with an A100 GPU under various CPU and GPU power caps reveals differing sensitivity patterns across applications. 
    }
    \label{fig:power_dual_cap}
    \vspace{-5pt}
\end{figure*}
We identify two key \emph{limitations} in existing studies: (1) Most existing works focus on improving energy efficiency through GPU power management alone, neglecting coordinated CPU-GPU power management for GPU applications. \cite{cpu-gpu} considers coordinated CPU-GPU power management, however it targets integrated CPU-GPU chips. Existing works fail to address CPU power management for GPU applications in CPU-GPU heterogeneous systems, yet our work demonstrates that coordinated CPU-GPU power management can achieve additional energy efficiency gains for GPU applications (see \S \ref{motivation and challenge}). (2) It is impractical for a system to conduct extensive offline profiling of every application to characterize their performance–power relationships. To support hierarchical power management, this limitation necessitates online performance profiling and prediction. However, existing hybrid and online-only approaches are unsuitable for several reasons. First, many hybrid approaches require prior knowledge of workloads or extensive offline profiling. Second, online-only approaches require lightweight models to minimize overhead, such as the linear regression model used in \cite{online-2} for GPU workload performance prediction, our experimental results demonstrate that linear regression models perform poorly in predicting the complex interplay between CPU-GPU power caps and workload performance (detailed results in \S \ref{section:sensitivity analysis}).

To bridge these gaps, we present OPEN, a hybrid approach which offers coordinated CPU-GPU power management with online prediction for CPU-GPU heterogeneous systems. The goal of OPEN is to maximize the energy efficiency while keep the application performance loss within a threshold. Upon an unseen application's arrival to a HPC system for execution, predicting application performance under wide range of CPU-GPU power settings is challenging due to the complex interplay between power caps and performance across diverse settings, while adapting to applications that transition between CPU-intensive and GPU-intensive execution phases. Hence, OPEN incorporates three primary techniques to tackle these challenges. First, OPEN build an offline performance predictor which is able to predict application performance given several hardware performance counters under various power settings. Second, since profiling all possible power settings for an unseen application online is infeasible, unlike previous hybrid approaches, OPEN doesn't require prior knowledge of unseen applications; instead, OPEN leverages \textit{collaborative filtering} with the offline trained performance predictor to estimate performance across a wide range of settings in an online manner. Third, OPEN provides a lightweight method for detecting phase transitions. Lastly, OPEN employs an energy-efficient power setting selection policy that ensures performance while maximizing energy efficiency, leveraging the prediction results.


We evaluate OPEN using a suite of GPU benchmarks and applications, including widely used GPU benchmarks from Altis \cite{hu2020altis, altis-sycl}, ECP proxy applications \cite{ecp}, three real-world HPC applications (LAMMPS, NAMD, GROMACS), three ML training workloads (UNet, ResNet50, and BERT) from the MLPerf benchmark \cite{farrell2021mlperf}, and HeCBench \cite{hec}. OPEN is compared against four power capping strategies: No-Power-Cap, GPU-Cap-Only, CPU-Cap-Only, and Oracle. Experimental results show that OPEN can predicts application performance under various CPU-GPU power cap settings, achieving up to 98\% prediction accuracy and 95\% on average. In terms of energy efficiency, OPEN delivers a 9.4\% improvement with only a 2.3\% performance loss. Furthermore, OPEN consistently outperforms No-Power-Cap, GPU-Cap-Only, CPU-Cap-Only, and achieves performance comparable to Oracle in many cases.
Overall, our work offers the following key contributions:

\begin{itemize}
    \item We analyze the impact of both CPU and GPU power capping on GPU applications running on heterogeneous systems, examining their performance and power characteristics to motivate the necessity and energy-saving potential of our approach (\S\ref{motivation and challenge}).
    \item We design OPEN, a coordinated CPU-GPU power management framework that leverages collaborative filtering to predict unseen application performance under various power settings in an online manner (\S\ref{OPEN design}).
    \item We extensively evaluate OPEN on different heterogeneous systems across various HPC benchmarks/application, and ML training workloads. (\S\ref{results}).
    \item OPEN will be released as open-source software on GitHub [link] to foster community-driven enhancements.
\end{itemize}

\section{Motivation and Challenges} \label{motivation and challenge}
Most existing studies emphasize GPU-side power management for GPU applications, often neglecting the critical role of CPU-side power control. To investigate the potential benefit from CPU power management for GPU applications, we perform extensive performance-power profiling across a diverse set of GPU applications and benchmarks. Our experiments reveal that applications can be broadly categorized into four groups based on their sensitivity to power capping: GPU-power-cap sensitive, CPU-power-cap sensitive, sensitive to both, and insensitive to both. In Figure~\ref{fig:power_dual_cap}, we include results for these four types of  applications and benchmarks. The results reveal that different applications exhibit distinct sensitivities to CPU and GPU power limitations (the details of these app are presented in Table \ref{tab:train_eval_apps}). For example, GROMACS demonstrates high sensitivity to CPU power capping while remaining largely unaffected by GPU power constraints. In contrast, BERT training shows significant sensitivity to GPU power capping but exhibits minimal response to CPU power variations. Additionally, UNet training and miniGAN demonstrate sensitivity to both CPU and GPU power constraints. Furthermore, these two applications exhibit similar power-performance patterns, suggesting that extensive offline profiling of applications may be unnecessary. Instead, during the online phase, one can search for similar power-performance patterns to predict an unseen application's characteristics without extensive offline profiling. 

\begin{figure}
    \centering
    \subfloat[miniGAN]{%
        \includegraphics[width=0.47\linewidth]{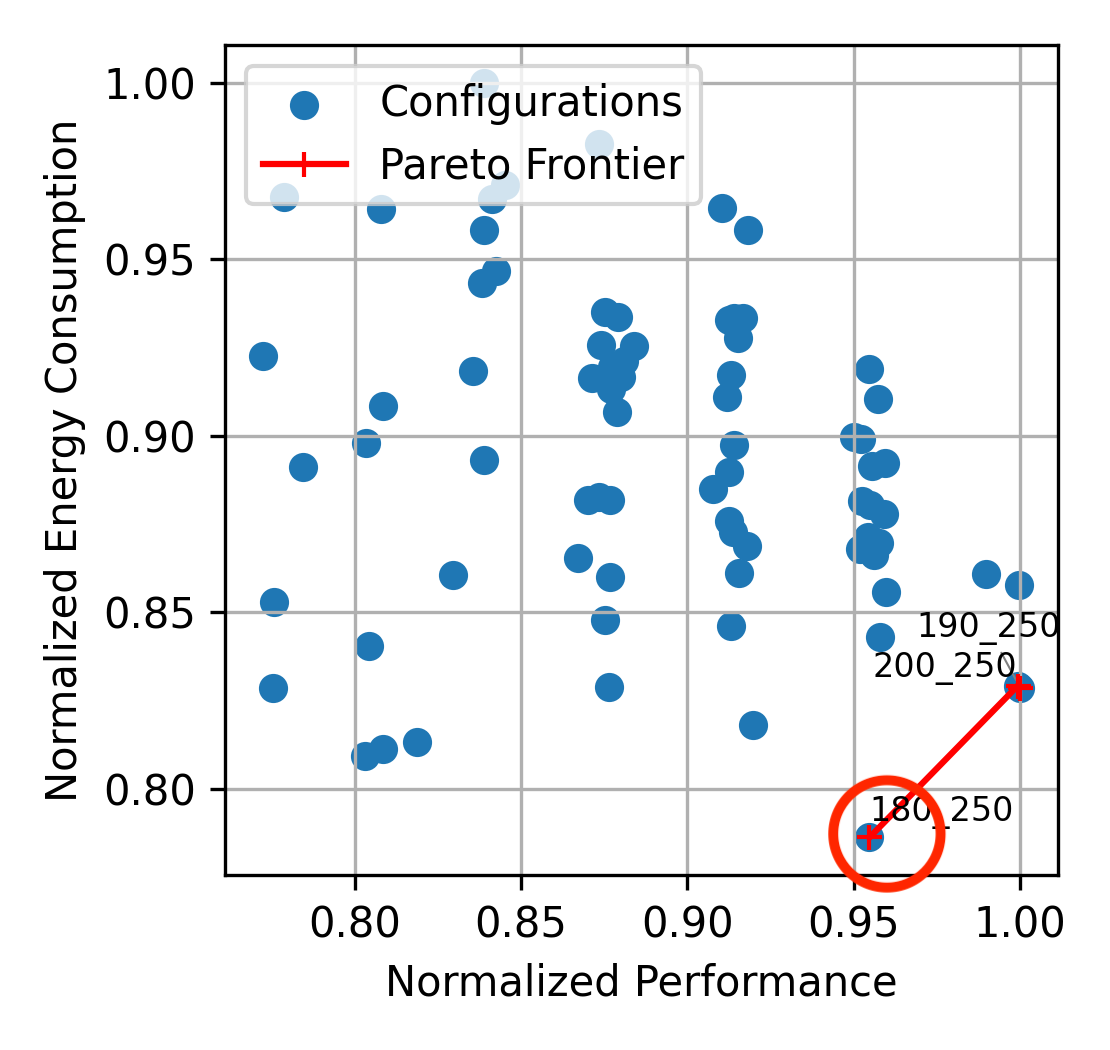}
        \label{fig:pareto_minigan}}
    \hspace{0.01\linewidth}
    \subfloat[UNet training]{%
        \includegraphics[width=0.47\linewidth]{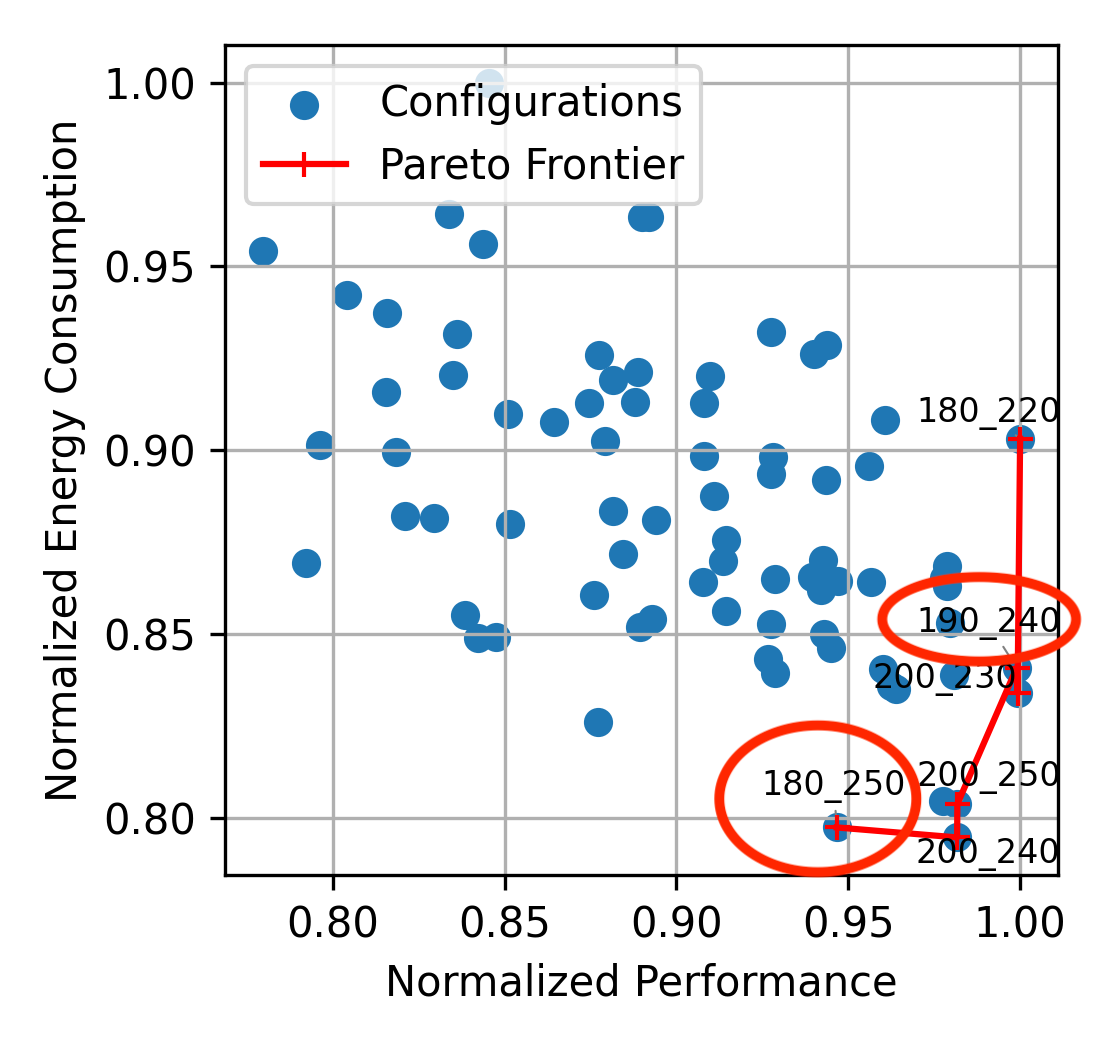}
        \label{fig:pareto_bert}}
    \caption{Pareto frontiers of applications between normalized performance and normalized energy consumption.}
    \label{fig:pareto_fig}
    \vspace{-5pt}
\end{figure}

In addition, we evaluated miniGAN and UNet training across a wide range of distinct CPU and GPU power cap settings. Figure~\ref{fig:pareto_fig} illustrates the Pareto frontier between normalized energy consumption and normalized performance. Notably, several settings on the frontier (highlighted with red circles) involve CPU power capping, indicating that CPU power management can provide additional energy efficiency opportunities. These observations demonstrate that applying CPU power caps can unlock significant potential for energy optimization---a gap that existing research has yet to address. Additionally, the proposed approach must account for the wide variety of diverse GPU applications, requiring the ability to quickly identify an unseen application's performance characteristics upon its arrival to the system for execution.

While the results above underscore the importance of coordinating CPU and GPU power capping across a wide range of applications, determining the optimal power caps for each component remains a significant challenge, especially considering the large number of applications deployed in real-world systems. In particular, we identify the following key challenges:

\begin{enumerate}

    \item \emph{Complex interplay between CPU-GPU power caps and application performance.} Coordinated CPU and GPU power capping introduces complex interactions affecting application performance, as shown in Fig \ref{fig:power_dual_cap}. Thus, the proposed approach must consider multiple performance metrics on both CPU and GPU. 
    
    \item \emph{Wide variety of applications and settings.} The number of applications and power capping settings is vast, making it impractical to test every combination for an unseen application. The approach must instead leverage limited online profiling to estimate performance under different power settings.

    \item \emph{Dynamic phase changing.} GPU workloads can begin with a long CPU-intensive phase for data preparation before shifting to GPU computation (Figure \ref{fig:power_pattern}). To manage power effectively, the approach must detect phase transitions timely and apply lightweight profiling that adapts to changing demands with minimal overhead.
\end{enumerate}

\section{OPEN Design} \label{OPEN design}
\begin{figure}
    \centering
    \includegraphics[width=1\linewidth]{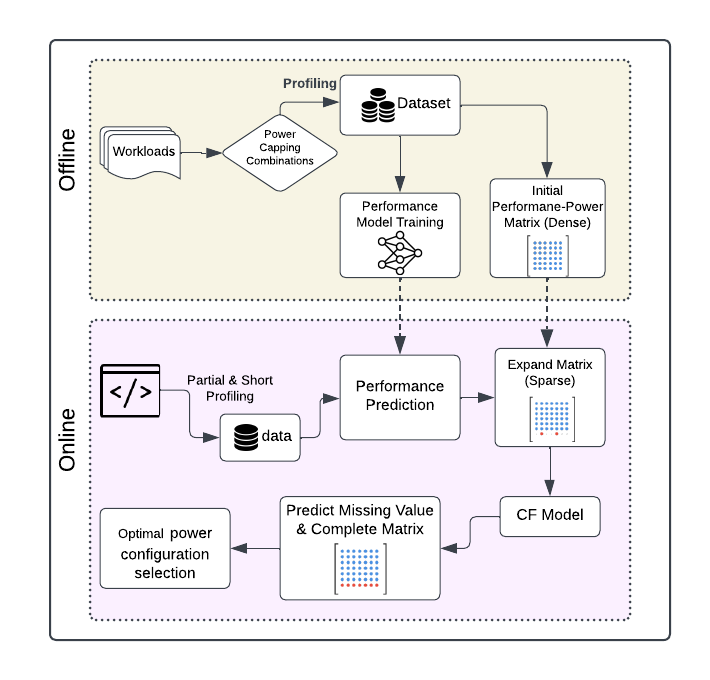}
    \caption{The OPEN framework consists of offline and online phases. The offline phase involves building a performance predictor and constructing an initial dense matrix. In the online phase, OPEN performs lightweight online profiling, and leverages the performance predictor with collaborative filtering to make performance prediction.}
    \label{fig:OPEN}
    \vspace{-5pt}
\end{figure}

Figure \ref{fig:OPEN} presents the architecture of OPEN, which consists of two main phases: offline profiling/modeling and online profiling/prediction. The following sections discuss the detailed design of OPEN.

\subsection{Offline Profiling and Modeling}

\subsubsection{Feature Selection} \label{feature_selection}

\begin{figure}
    \centering
    \includegraphics[width=1\linewidth]{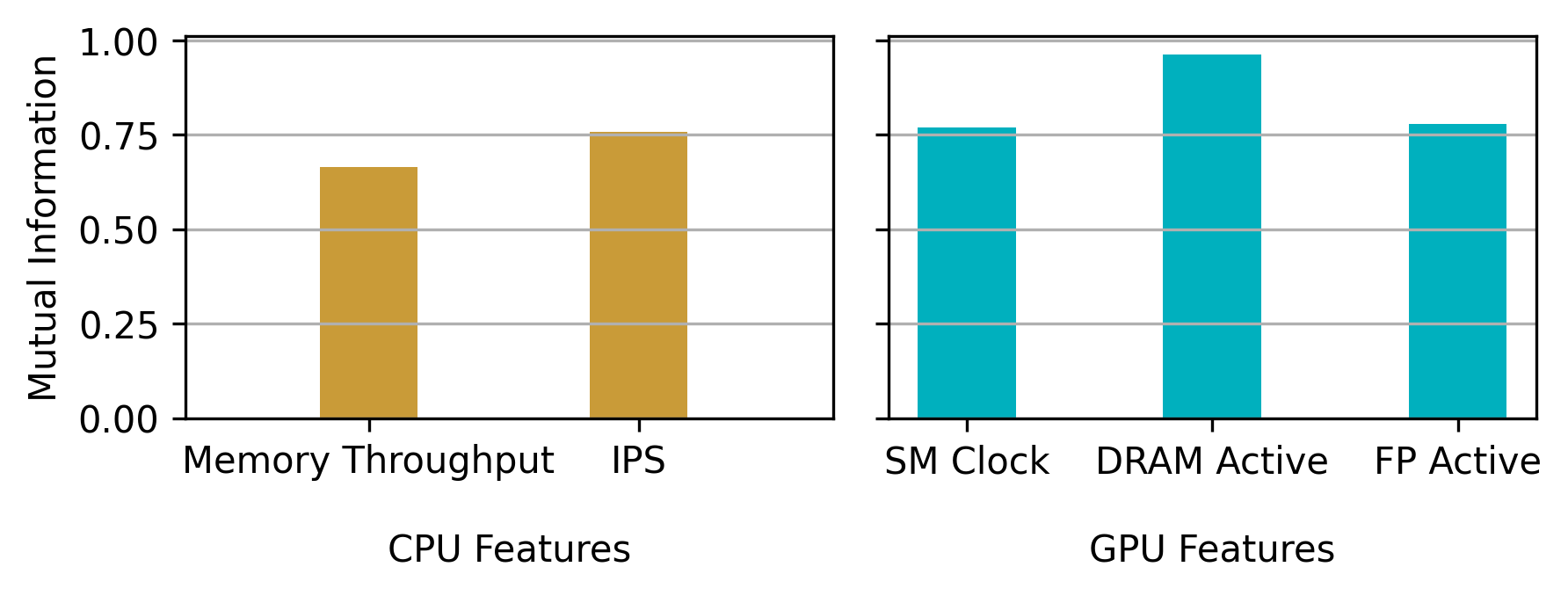}
    \caption{Mutual information scores between selected CPU and GPU performance metrics and application performance.}
    \label{fig:feature_importance}
\end{figure}

Determining appropriate performance metrics is critical to predict application performance under different power settings. To identify the informative features, we conduct a mutual information analysis on both CPU and GPU performance counters, considering GPU applications listed in Table \ref{tab:train_eval_apps}. Following the findings of Ali et al. \cite{ali2023performance}, we include SM\_Clock, FP\_Active, and DRAM\_Active to capture key GPU behaviors under different GPU power cap settings. On the CPU side, we incorporate Instructions Per Second (IPS) and DRAM throughput, to represent compute and memory activities, respectively. Figure \ref{fig:feature_importance} presents the mutual information scores between these features and application performance. The left subplot shows CPU metrics, where memory throughput is the most informative (MI = 0.88), followed by IPS (MI = 0.55). The right subplot shows GPU metrics, with DRAM Active yielding the highest score (MI = 0.96), followed by FP Active (MI = 0.78) and SM Clock (MI = 0.77). These results indicate that both CPU and GPU features contribute significantly to performance prediction, with memory-related counters on both sides demonstrating the strongest influence. Table \ref{tab:perf_events} summarizes the selected performance counters used as input features in our prediction models in \S \ref{section:mlp_model}.

\begin{table}[h]
    \centering
    \caption{Performance Metrics Descriptions}
    \begin{tabular}{ll}
        \toprule
        \textbf{Performance Event} & \textbf{Description} \\
        \midrule
        Instruction Per Second (IPS) & The number of retired instructions \\
        & per second. \\
        DRAM Memory-Throughput & Read \& Write Memory-Throughput \\
        SM\_Active & Percentage of active SM \\
        FP\_Active & Percentage of GPU's FP execution \\ 
        &  units are processing instructions \\
        DRAM\_Active & GPU's DRAM is actively \\
        & servicing memory requests \\
        \bottomrule
    \end{tabular}
    \label{tab:perf_events}
\end{table}

\subsubsection{Performance Prediction Model}
\label{section:mlp_model}

Some previous works have utilized linear regression models \cite{model-5} and traditional non-linear analytical approaches \cite{model-1, model-2, model-3, model-4, model-6} such as polynomial regression methods to predict application performance in computing systems. However, the complexity of predicting performance has increased dramatically when involving the intricate interactions between CPU and GPU performance counters under various power capping constraints, where traditional analytical models struggle to capture the non-linear dependencies and complex feature interactions inherent in modern heterogeneous computing environments, and some works \cite{nn_better_1, nn_better_2} argues that neural network models can effectively capture complicated non-liner relation. We compare Multi-Layer Perceptron model, linear models (Linear Regression, Ridge Regression) and non-linear analytical models (Random Forest, Gradient Boosting, K-Nearest Neighbors), and the experiments result show that MLP model performs better. 


Our performance prediction model takes as input a vector \(\mathbf{X} \in \mathbb{R}^d\), where \(d\) is the number of selected performance features relevant to the application type. The input vector \(\mathbf{X}\) consists of seven features: CPU power cap, GPU power cap, Memory Throughput, IPS, Streaming Multiprocessor (SM) Clock Frequency, DRAM Active rate, and Floating Point (FP) Active rate. These features collectively capture the compute and memory behaviors of both CPU and GPU components under power constraints. Eventually this model predicts the normalized performance for an application running under a power setting (CPU and GPU power caps).



\subsubsection{Putting it all together}

During the offline profiling and modeling phase, a performance prediction model is constructed to characterize application performance across different CPU and GPU power settings, based on profiling a limited number of representative GPU benchmarks (see Table \ref{tab:train_eval_apps}). Specifically, we perform detailed offline profiling of these benchmarks by collecting hardware counters and runtime measurements under various power settings. The collected data serves two purposes: (1) to build a performance prediction model that estimates an application's normalized performance relative to the no-power-cap case based on hardware counter readings, and (2) to construct an initial performance-power matrix, $P_{m,n}$. 

OPEN construct an initial dense matrix, presented in Matrix \ref{eq:performance_matrix}, where each row represents a power setting (e.g., 200W CPU and 250W GPU power caps) and each column corresponds to an application. For instance, P\textsubscript{m,n} represents the performance (normalized runtime) for application \textit{m} under power setting \textit{n}. This matrix serves as the foundation for the \textit{collaborative filtering} \ref{NCF_model} to learn from known data in the online phase.

\begin{equation}
    \label{eq:performance_matrix}
    P_{m,n} = 
    \begin{pmatrix}
    P_{1,1} & P_{1,2} & \cdots & P_{1,n} \\
    P_{2,1} & P_{2,2} & \cdots & P_{2,n} \\
    \vdots  & \vdots  & \ddots & \vdots  \\
    P_{m,1} & P_{m,2} & \cdots & P_{m,n}
    \end{pmatrix}
\end{equation}

\subsection{Online Profiling and Prediction}


\subsubsection{Collaborative Filtering Model} \label{NCF_model}
Collaborative Filtering (CF) is a foundational technique in recommender systems, widely adopted following its popularization during the Netflix Prize competition \cite{cf_netflex}. The key idea is to predict missing entries in a sparse user–item interaction matrix by leveraging patterns learned from known data. Several CF models have been developed to solve this prediction task, such as K-Nearest Neighbors (KNN) \cite{cf_knn}, Singular Value Decomposition (SVD) \cite{cf_svd}, autoencoders \cite{cf_autoencoder}, and Neural Collaborative Filtering (NCF) \cite{neural_CF}. NCF replaces linear dot-product interactions with neural networks that learn non-linear patterns. User and item embeddings are passed through a multi-layer perceptron to predict scores, enabling the model to capture complex interaction dynamics beyond linear correlations. 

Therefore, to address the challenge (2) in \S \ref{motivation and challenge}, we adopt NCF to estimate the workload performance using a few sampled data points. Specifically, We model the problem of performance prediction under varying power capping settings as a matrix completion task. The performance matrix is presented in Equation 1, where each column corresponds to an application and each row corresponds to a specific power capping setting. The element \( P_{i,j} \in \mathbb{R} \) represents the observed or predicted performance of application \( i \) under setting \( j \).

The NCF model driven by this work \cite{neural_CF}. The input to the model consists of a Power Cap Pair Index \( j \in \{1, \dots, n\} \) and an Application Index \( i \in \{1, \dots, m\} \). These indices are mapped to dense latent vectors through embedding layers, producing compact representations that capture the underlying characteristics of each application and setting. The resulting embeddings are then concatenated to form a joint feature vector that represents the interaction between a specific application and a given power cap setting. This joint vector is passed through a multi-layer perceptron (MLP) with non-linear activation functions, such as SELU, which enables the network to model complex, high-order interactions. The final output of the MLP is a scalar value that represents the predicted performance of the application under the specified setting. By learning to generalize from observed performance data, the NCF model is capable of estimating unobserved entries in the performance matrix, even for previously unseen combinations of applications and power cap settings. This modeling approach allows for flexible and scalable performance prediction across a wide range of CPU-GPU power settings.


\begin{figure}
    \centering
    \subfloat[GROMACS]{%
        \includegraphics[width=0.51\linewidth]{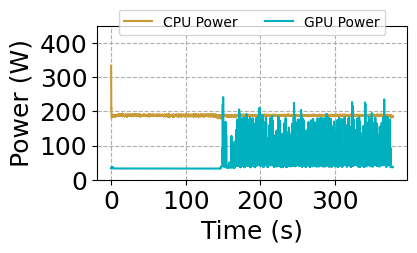}
        \label{fig:gromacs_power}}
    \hspace{0.01\linewidth}
    \subfloat[BERT Training]{%
        \includegraphics[width=0.41\linewidth]{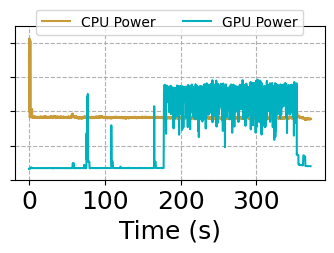}
        \label{fig:bert_power}}
    \caption{Applications with long CPU phases.}
    \label{fig:power_pattern}
\end{figure}

\subsubsection{Online Performance Prediction}

The online phase consists of two steps: online profiling and online prediction. First, OPEN performs lightweight online profiling. For each unseen application submitted to the system for execution, OPEN runs it for a short interval under several power settings, collecting performance counters during each run. These counters are then fed into the offline performance predictor to predict normalized performance values, which are used to expand the dense matrix (Matrix \ref{eq:performance_matrix}) to a sparse matrix. To impute these missing values, which are application performance under other power settings, in the sparse matrix, we employ collaborative filtering. In this work, we adopt a Neural Collaborative Filtering (NCF) model, which leverages similarities in performance patterns between previously profiled applications and the new application. The details of collaborative filtering model selection is discussed in Section \ref{NCF_model}.

\begin{algorithm}[H]
\caption{Phase Transition Detection}
\label{alg:phase_transition}
\begin{algorithmic}[1]
\State \textbf{Input:} Power sampling interval $\delta = 0.2$ seconds, window size $W = 5$ seconds, threshold $P_{\text{th}} = 60$W
\State \textbf{Initialize:} Circular buffer $Q$ with capacity $N = W / \delta = 25$
\While{application is running}
    \State $p \gets$ current GPU power reading
    \State Append $p$ to $Q$ (remove oldest if full)
    \If{length of $Q = N$}
        \State Divide $Q$ into 5 consecutive 1-second intervals
        \State $flag \gets$ \textbf{true}
        \For{each 1-second interval $I$ in $Q$}
            \If{any power value in $I$ is less than $P_{\text{th}}$}
                \State $flag \gets$ \textbf{false}
                \State \textbf{break}
            \EndIf
        \EndFor
        \If{$flag = \textbf{true}$}
            \State \textbf{return} Phase transition detected
        \EndIf
    \EndIf
\EndWhile
\end{algorithmic}
\end{algorithm}

However, sometimes short-interval profiling is insufficient for GPU applications that exhibit long CPU-side phases (e.g., data preparation) before launching GPU workloads. Examples include GROMACS and BERT training, as illustrated in Fig. \ref{fig:power_pattern}, which shows their power consumption patterns. In such cases, brief profiling fails to capture the performance and power behavior of the GPU-intensive phases. To address the challenge (3) in \S \ref{motivation and challenge}, we propose a multi-phase profiling solution. The first step involves detecting the transition from CPU to GPU activity. We monitor GPU power consumption periodically and maintain a sliding window recording the recent history power data. A phase transition is identified when the GPU power exceeds a predefined threshold lasting for some time detailed in Algorithm 1. Once a transition is detected, we initiate a second round of online profiling, running a few power settings with each for one seconds. This duration is sufficient because GPU power patterns are highly repetitive in representative HPC applications such as GROMACS and deep learning workloads such as BERT.


\begin{algorithm} 
\caption{Energy Efficient Power Cap Selection}
\textbf{Input:} performance matrix $M$, CPU caps $C$, GPU caps $G$, baseline CPU cap $C_b$, baseline GPU cap $G_b$, performance loss threshold $\gamma$\\
\textbf{Output:} selected power cap pair $(c^*, g^*)$

\begin{algorithmic}[1]
\State $p_\text{base} \gets M[(C_b, G_b)]$
\State $E_\text{base} \gets (C_b + G_b) \cdot 1$ \Comment{Normalized upper bound energy consumption of baseline}
\State $P \gets \emptyset$ \Comment{Valid power cap options}

\For{each $(c, g) \in C \times G$}
    \State $p \gets M[(c, g)]$ \Comment{Predicted normalized performance}
    \State $\text{pred\_loss} \gets 1 - \frac{p}{p_\text{base}}$
    \If{$\text{pred\_loss} \leq \gamma$}
        \State $E_\text{pred} \gets (c + g) \cdot \frac{1}{p}$
        \State $\text{pred\_saving} \gets \frac{E_\text{base} - E_\text{pred}}{E_\text{base}}$
        \State Add $(c, g, \text{pred\_saving})$ to $P$
    \EndIf
\EndFor
\State Select $(c^*, g^*) \in P$ with maximum $\text{pred\_saving}$
\State \textbf{return} $(c^*, g^*)$
\end{algorithmic}
\end{algorithm}

\subsubsection{Optimal Power setting Selection} \label{power cap selection}
\begin{table}[h]
    \centering
    \caption{Applications Used for Training and Evaluation}
    \begin{tabular}{l l l}
        \toprule
        \textbf{Suite} & \textbf{Application} & \textbf{Size/Dataset/Input} \\
        \midrule
        \multicolumn{3}{c}{\textbf{Training App}} \\
        \midrule
        Altis & gemm & \multicolumn{1}{c}{-s 4} \\
        & gups & \multicolumn{1}{c}{-s 4} \\
        & maxflops & \multicolumn{1}{c}{/} \\
        & bfs & \multicolumn{1}{c}{-s 4} \\
        & particlefilter\_float & \multicolumn{1}{c}{-s 4} \\
        & cfd\_double & \multicolumn{1}{c}{-s 4} \\
        & particlefilter\_naive & \multicolumn{1}{c}{-s 4} \\
        & raytracing & \multicolumn{1}{c}{-s 4} \\
        & fdtd2d & \multicolumn{1}{c}{-s 4} \\
        & nw & \multicolumn{1}{c}{-s 4} \\
        \midrule
        \multicolumn{3}{c}{\textbf{Evaluation App}} \\
        \midrule
        Altis & cfd & \multicolumn{1}{c}{-s 4} \\
        & lavamd & \multicolumn{1}{c}{-s 4} \\
        & sort & \multicolumn{1}{c}{-s 3} \\
        \midrule
        MLPerf Training & UNet & carvana-image\\
        & BERT & Wiki\\
        & ResNet50 & ImageNet \\
        \midrule
        Real HPC App & GROMACS & Steepest descent,\\
        & & emtol=1000, PBC=xyz\\
        & & PME, rc=1.0 nm, \\
        & LAMMPS & 3D LJ melt \\
        & & 6.25M atoms, 500 steps, \\
        & & cutoff 2.5 \\
        & NAMD & stmv\_nve\_cuda.namd \\
        & miniGAN & bird, 2048 images, \\
        & & 3 channels, 64×64, \\
        & & dim-mode 3\\
        \midrule
        ECP Proxy & sw4lite & ps2.in \\
        & XSBench & -s large\\
        & Laghos & box01\_hex.mesh \\
        \midrule
        HeCBench & kalman & 10000 10000 10000\\
        & stencil3d & 1100 \\
        & extrema & 5000 \\
        & knn & 2000 \\
        & dropout & 100000000 \\
        & aobench & 50000 \\
        & zoom & 64 32 512 512\\
        & convolution3D & 32 64 128 56 56 3\\
        & softmax & 10000 100000 \\
        & chacha20 & 1000000 \\
        & zmddft & 10000 \\
        \bottomrule
    \end{tabular}
    \label{tab:train_eval_apps}
\end{table}
To translate OPEN's performance predictions into actionable power management strategies, we define a power setting selection policy shown in Algorithm 2 to identify the optimal CPU–GPU power cap setting. When a user submits an application to the system, they are encouraged to provide a performance loss threshold to ensure performance degradation remains within predefined limits (e.g., 5\%) while maximizing energy savings. OPEN then identifies the optimal CPU–GPU power cap setting within the performance loss threshold.

In algorithm 2, we begin by defining the baseline power cap pair as $(C_b, G_b)$: the CPU and GPU caps (e.g., 200W CPU, 250W GPU). This defines the normalized upper bound for energy consumption as $E_{\text{base}} = (C_b + G_b) \cdot 1$. For each candidate power cap pair $(c, g)$, we retrieve the predicted normalized performance $\hat{p}_{(c,g)}$ from the performance matrix predicted by the MLP and NCF models. We then compute the predicted performance loss as $\text{pred\_loss}_{(c,g)} = 1 - \hat{p}_{(c,g)} / \hat{p}_{(C_b, G_b)}$. Only settings where $\text{pred\_loss}_{(c,g)} \leq \gamma$ are considered valid (e.g., $\gamma = 0.05$). For each valid setting, we calculate an upper bound of the predicted energy consumption as $E_{(c,g)} = (c + g) / \hat{p}_{(c,g)}$. From this, the predicted percentage of energy saving is estimated as $\text{Saving}_{(c,g)} = (E_{\text{base}} - E_{(c,g)}) / E_{\text{base}}$. The energy efficient setting $(c^*, g^*)$ is selected as the pair that yields the maximum predicted energy saving, i.e., $(c^*, g^*) = \arg\max_{(c,g)} \text{Saving}_{(c,g)}$.

This policy enables lightweight and effective decision-making without requiring exhaustive runtime profiling. OPEN thereby supports efficient power-aware scheduling across diverse GPU applications.

\section{Experimental setting} \label{experiment setup}


We select representative heterogeneous applications and benchmarks for performance modeling and evaluation in Table \ref{tab:train_eval_apps} respectively, and we present the details of training and evaluation applications below.

\begin{figure*}
    \centering
    \includegraphics[width=1\linewidth]{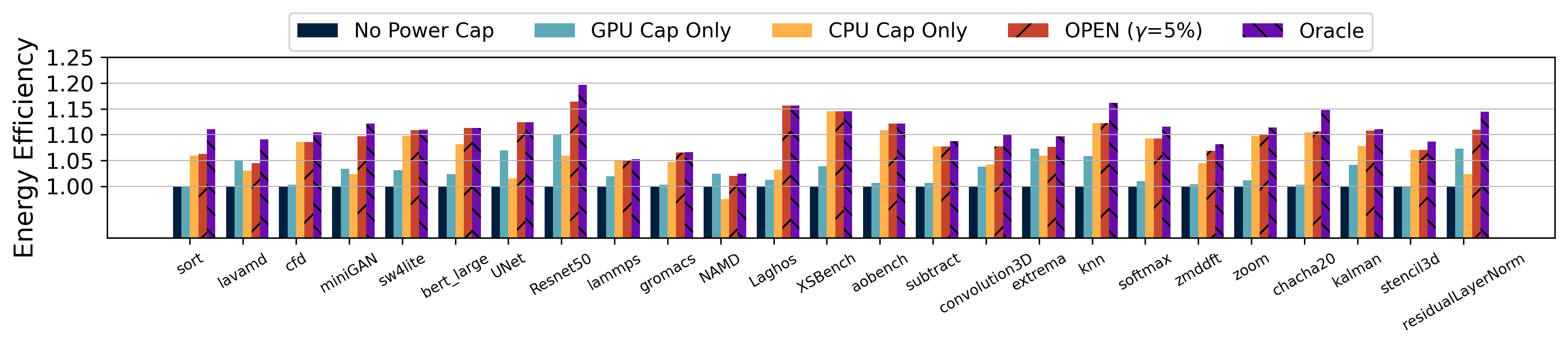}
    \caption{Normalized application energy efficiency on A100 with different power capping strategies.}
    \label{fig:energy_efficiency_a100}
\end{figure*}

\begin{figure*}
    \centering
    \includegraphics[width=1\linewidth]{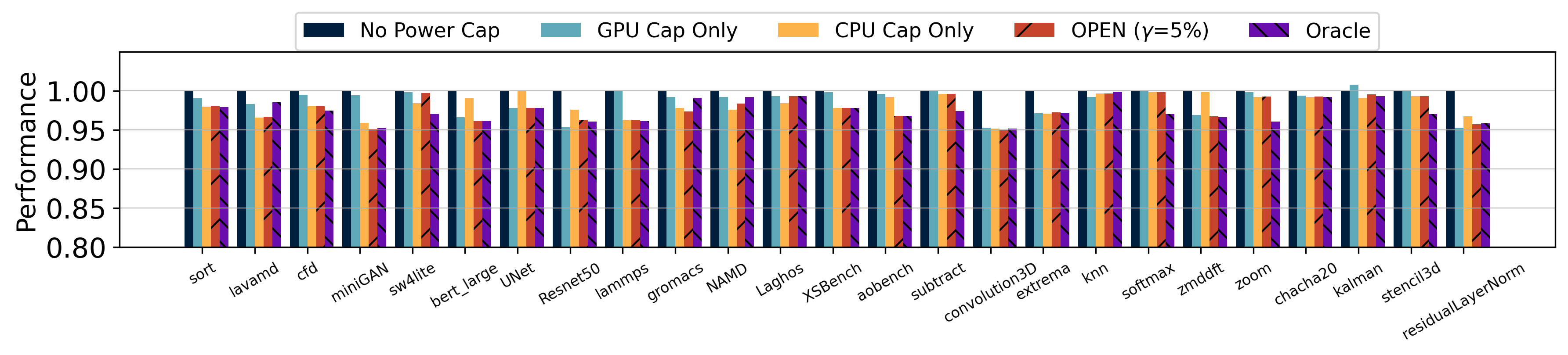}
    \caption{Normalized application performance on A100 with different power capping strategies.}
    \label{fig:performance_a100}
\end{figure*}

\textbf{\emph{Training Applications.}} We use the Altis GPU benchmark suite \cite{hu2020altis}, designed for evaluating heterogeneous computing systems with GPU and accelerator workloads across multiple parallel programming models such as CUDA and OpenCL. From Altis, we select 10 applications from Levels~1 and~2, which include fundamental parallel algorithms widely used in high-performance computing and real-world workloads. These applications exhibit diverse sensitivities to CPU and GPU power capping, making them well suited for training the performance prediction model.

\textbf{\emph{Evaluation Applications.}} For evaluation, we select: four applications from the ECP proxy app suite \cite{ecp}, which represents the computational patterns of full-scale scientific applications for exascale system testing; three neural network training workloads from MLPerf benchmarks \cite{farrell2021mlperf}, an industry-standard suite for measuring machine learning performance across hardware platforms; three molecular simulation applications \cite{van2005gromacs, atomic2013lammps, NAMD} representative of real GPU-enabled HPC workloads; and 11 applications from HecBench \cite{hec}, a benchmark suite for evaluating HPC systems using representative workloads from diverse scientific domains; and 3 apps from Altis. Based on their responses to power capping, these benchmarks and applications fall into four categories: GPU-power-cap sensitive, CPU-power-cap sensitive, sensitive to both, and insensitive to both, making them robust for evaluation.

Two distinct heterogeneous computing systems are used in our experiments:

\begin{itemize}
    \item \textbf{Intel+A100}: A Chameleon Cloud \cite{keahey2020lessons} system featuring two Intel(R) Xeon(R) Platinum 8380 processors paired with a single NVIDIA A100-40GB GPU. 
    \item \textbf{Intel+A30}: A Chameleon Cloud \cite{keahey2020lessons} system featuring two Intel(R) Xeon(R) Platinum 8380 processors paired with a single NVIDIA A30-25GB GPU.
\end{itemize}

We compare OPEN against four power capping strategies: No-Power-Cap, GPU-Cap-Only, CPU-Cap-Only, and Oracle. We evaluate each method using two key metrics:

\begin{itemize}
    \item \textbf{Performance Loss.} Defined as the percentage increase in execution time compared to the baseline, capturing the performance impact of power capping.
    \item \textbf{Energy Efficiency \cite{zhang2024improving}:} Defined as the normalized performance divided by normalized average power. Power consumption includes CPU package, DRAM, and GPU board power, compared to the baseline. GPU board includes GPU cores, GPU memory, and GPU onboard components (VRM, fans, PCIe logic). 
\end{itemize}


\section{Results} \label{results}
Our experimental results evaluate OPEN across GPU applications on multiple heterogeneous systems, including A100, and A30 GPUs. Each experiment was repeated at least five times to account for performance variability and system-level noise. Outliers were removed, and the average of the remaining results was used to ensure reliability and robustness.  


\subsection{End-to-End Performance} \label{section:energy saving}
OPEN provides performance predictions across a wide range of coordinated CPU-GPU power cap settings. Building on this capability, our goal is to translate predictive accuracy into tangible energy savings by selecting power cap settings that optimize energy efficiency. In this section, we evaluate the effectiveness of the Energy-Efficient Power Cap Selection algorithm introduced in Section \ref{power cap selection}. Specifically, we evaluate OPEN with other power capping policy using \textit{energy-efficiency} and \textit{performance loss} defined in Section \ref{experiment setup}. Figure \ref{fig:energy_efficiency_a100}, and Figure \ref{fig:performance_a100} shows the \textit{energy efficiency} and \textit{performance loss} achieved under five power capping policies: No-Power-Cap, GPU-Cap-Only, CPU-Cap-Only, Oracle, and OPEN. The energy efficiency of the No-Power-Cap policy is used as the baseline and is normalized to 1. To determine the optimal power capping setting for Oracle, we fully execute each application under tested power capping settings. In the presented evaluation, we enforce a performance loss threshold of 5\% for the energy-efficiency power setting selection algorithm, however, later in this section we will present the trade-offs between performance loss threshold and energy-efficiency. 

We highlight several observations from the results. First, OPEN consistently outperforms No-Power-Cap, achieving an average improvement of 9.4\% in energy efficiency while incurring only a 2.3\% average performance loss, with the performance loss threshold set to 5\%. Specifically, for ResNet50 training, OPEN achieves up to a 17\% improvement in energy efficiency with only ~3.5\% performance loss compared to the No-Power-Cap baseline. Second, while GPU-Cap-Only and CPU-Cap-Only policies also improve energy efficiency for most applications compared to the No-Power-Cap baseline, OPEN is able to outperform or match their performance. For example, in \textit{bert\_large} training, GPU-Cap-Only achieves only a 2.5\% improvement in energy efficiency because the application is sensitive to GPU power capping, as shown in Figure \ref{fig:power_dual_cap}. However, the coordinated CPU–GPU power capping introduced by OPEN surpasses this limit. Since \textit{bert\_large} is not sensitive to CPU power capping, OPEN increases energy efficiency to 12.5\% (from 2.5\%) while adding only 0.5\% extra performance loss compared to GPU-Cap-Only. Third, for some applications, such as \textit{lavamd}, \textit{cfd}, and \textit{XSBench}, OPEN delivers higher energy efficiency than GPU-Cap-Only, albeit with slightly greater performance degradation. For instance, for \textit{XSBench}, OPEN achieves 13\% higher energy efficiency than GPU-Cap-Only at the cost of an additional 3\% performance loss. This trade-off is expected, as more aggressive power capping through coordinated CPU–GPU power management introduces a balance between energy savings and performance. Fourth, OPEN achieves performance comparable to Oracle for many applications, such as \textit{sw4lite}, \textit{bert\_large}, and \textit{UNet}.

Table \ref{tab:energy_efficiency} summarizes the energy efficiency gains and performance losses for OPEN compared to No-Power-Cap on both A100 and A30 systems. We make several observations. First, for both systems, OPEN delivers noticeable improvements in energy efficiency while incurring negligible performance loss. For example, on the A100 system, OPEN improves energy efficiency by an average of 9.4\% with only a 2.3\% performance loss, given a performance loss threshold of 5\%. Second, for both systems, loosening the performance loss threshold allows for further improvements in energy efficiency; however, this comes at the cost of increased performance loss. This trade-off provides valuable flexibility for cluster-level decision-making, enabling administrators to balance energy savings and performance according to operational priorities. In summary, by coordinately managing both CPU and GPU power settings, OPEN provides enhanced energy efficiency while maintaining comparable performance to GPU-Cap-Only and CPU-Cap-Only policies.

\begin{table}[h]
    \centering
    \caption{Average Energy efficiency improvement and performance loss under different performance loss thresholds by OPEN.}
    \label{tab:energy_efficiency}
    \begin{tabular}{lcc|cc}
    \toprule
    \textbf{Threshold} & \multicolumn{2}{c|}{\textbf{A100}} & \multicolumn{2}{c}{\textbf{A30}} \\
     & \textbf{EE Gain} & \textbf{Perf. Loss} & \textbf{EE Gain} & \textbf{Perf. Loss} \\
    \midrule
    $\gamma=5\%$ & 9.4\% & 2.3\% & 7.8\% & 1.2\% \\
    $\gamma=10\%$ & 10.5\% & 5.6\% & 8.9\% & 4.8\% \\
    $\gamma=15\%$ & 11.3\% & 6.8\% & 9.6\% & 6.1\% \\
    \bottomrule
    \end{tabular}
\end{table}

\section{Related Work}
As GPU workloads have become increasingly dominant in HPC systems \cite{zhao2023power}, power capping and frequency scaling have emerged as widely adopted techniques to improve energy efficiency \cite{dutta2018gpu, fan2019predictable, wang2020gpgpu, guerreiro2019dvfs, wu2015gpgpu, ali2023performance, guerreiro2018gpgpu, dvfs}. Since different power caps and frequency settings can result in varying levels of application performance degradation, understanding and predicting application performance under diverse power and frequency settings is crucial for selecting energy-efficient settings.

Extensive research has focused on building offline models, including statistical and neural network approaches, to predict power consumption and performance \cite{offline-1, offline-2, offline-3, offline-4, offline-5, offline-6, offline-7, offline-8, offline-9, offline-10, offline-11}. However, these approaches suffer from several limitations. Some require detailed source code analysis and application modifications \cite{offline-4, offline-7}, limiting their practical applicability. Moreover, these works focus exclusively on offline modeling and prediction without proposing online prediction mechanisms or adaptive frequency scaling and power capping policies for real-time energy efficiency improvements.

To address these limitations, several approaches have been proposed that integrate offline workload profiling with online power and performance prediction, combined with adaptive frequency and power capping policies \cite{hybrid-1, hybrid-2, hybrid-3, hybrid-4, hybrid-5, hybrid-6, hybrid-7, hybrid-8, hybrid-9, hybrid-10, hybrid-11, hybrid-12, sim-1, sim-2, sim-3}. Some works \cite{hybrid-8, hybrid-5, hybrid-4} do not provide performance guarantees when applying frequency scaling or power capping. Several approaches \cite{hybrid-3, hybrid-9, hybrid-8, hybrid-10} assume prior knowledge of newly arriving workloads, including specific GPU kernels, and require extensive offline kernel profiling; and, some works \cite{sim-1, sim-2, sim-3} utilize online simulators for prediction but necessitate a complete execution pass to collect low-level hardware performance counters.

A few studies \cite{online-1, online-2} have proposed fully online frequency scaling policies with performance assurance. However, one approach \cite{online-1} is limited to specific workloads, such as iterative HPC applications, and fails to address phase-changing workloads. Another work \cite{online-2} addresses this limitation by proposing a multi-probing mechanism that periodically monitors and profiles workload performance to guide frequency scaling decisions. Notably, none of the aforementioned works consider coordinated CPU and GPU power capping for GPU workloads. Our analysis demonstrates that CPU power capping can improve energy efficiency for workloads that are insensitive to CPU power constraints, as shown in Figure \ref{fig:power_dual_cap}.

Collaborative filtering techniques \cite{cf}, widely adopted in recommendation systems to infer missing values based on inter-item or inter-user correlations, have also found applications in HPC and cloud computing. For instance, Paragon \cite{delimitrou2013paragon} employs collaborative filtering to predict workload performance and interference during colocation under different system settings. Similarly, Salaria et al. \cite{system_cf} apply collaborative filtering to forecast HPC workload performance across various systems. These studies demonstrate the potential of collaborative filtering for performance prediction in computing systems.
To address the limitations of prior work, we employ collaborative filtering to enable online workload performance prediction under diverse power cap settings without requiring extensive and detailed offline profiling.

\section{Conclusion}
This paper presents OPEN, an online performance prediction framework for heterogeneous systems operating under CPU and GPU power caps. By leveraging collaborative filtering for online performance prediction, OPEN significantly reduces profiling overhead while maintaining high prediction accuracy. Experimental results on diverse GPU platforms demonstrate that OPEN achieves up to 98.29\% prediction accuracy. These results highlight the effectiveness of OPEN in enabling scalable, low-overhead performance modeling and its potential for guiding energy-efficient decision making in real-world HPC environments.

\newpage
\bibliographystyle{ieeetr}
\bibliography{bib/IEEE.bib}

@String{Computing = "Computing" }

@String{Computer = "{IEEE} Computer" }

@String{Springer = "Springer-Verlag" }

@article{delimitrou2013paragon,
  title={Paragon: QoS-aware scheduling for heterogeneous datacenters},
  author={Delimitrou, Christina and Kozyrakis, Christos},
  journal={ACM SIGPLAN Notices},
  volume={48},
  number={4},
  pages={77--88},
  year={2013},
  publisher={ACM New York, NY, USA}
}

@inproceedings{hec,
  title={A benchmark suite for improving performance portability of the sycl programming model},
  author={Jin, Zheming and Vetter, Jeffrey S},
  booktitle={2023 IEEE International Symposium on Performance Analysis of Systems and Software (ISPASS)},
  pages={325--327},
  year={2023},
  organization={IEEE}
}

@inproceedings{system_cf,
  title={Predicting performance using collaborative filtering},
  author={Salaria, Shweta and Drozd, Aleksandr and Podobas, Artur and Matsuoka, Satoshi},
  booktitle={2018 IEEE International Conference on Cluster Computing (CLUSTER)},
  pages={504--514},
  year={2018},
  organization={IEEE}
}

@inproceedings{ali2023performance,
  title={Performance-aware energy-efficient GPU frequency selection using DNN-based models},
  author={Ali, Ghazanfar and Side, Mert and Bhalachandra, Sridutt and Wright, Nicholas J and Chen, Yong},
  booktitle={Proceedings of the 52nd International Conference on Parallel Processing},
  pages={433--442},
  year={2023}
}

@inproceedings{hu2020altis,
  title={Altis: Modernizing gpgpu benchmarks},
  author={Hu, Bodun and Rossbach, Christopher J},
  booktitle={2020 IEEE International Symposium on Performance Analysis of Systems and Software (ISPASS)},
  pages={1--11},
  year={2020},
  organization={IEEE}
}

@article{NAMD,
  title={Scalable molecular dynamics with NAMD},
  author={Phillips, James C and Braun, Rosemary and Wang, Wei and Gumbart, James and Tajkhorshid, Emad and Villa, Elizabeth and Chipot, Christophe and Skeel, Robert D and Kale, Laxmikant and Schulten, Klaus},
  journal={Journal of computational chemistry},
  volume={26},
  number={16},
  pages={1781--1802},
  year={2005},
  publisher={Wiley Online Library}
}

@inproceedings{guerreiro2018gpgpu,
  title={GPGPU power modeling for multi-domain voltage-frequency scaling},
  author={Guerreiro, Joao and Ilic, Aleksandar and Roma, Nuno and Tomas, Pedro},
  booktitle={2018 IEEE International Symposium on High Performance Computer Architecture (HPCA)},
  pages={789--800},
  year={2018},
  organization={IEEE}
}

@article{guerreiro2019dvfs,
  title={DVFS-aware application classification to improve GPGPUs energy efficiency},
  author={Guerreiro, Jo{\~a}o and Ilic, Aleksandar and Roma, Nuno and Tom{\'a}s, Pedro},
  journal={Parallel Computing},
  volume={83},
  pages={93--117},
  year={2019},
  publisher={Elsevier}
}

@inproceedings{dutta2018gpu,
  title={GPU power prediction via ensemble machine learning for DVFS space exploration},
  author={Dutta, Bishwajit and Adhinarayanan, Vignesh and Feng, Wu-chun},
  booktitle={Proceedings of the 15th ACM International Conference on Computing Frontiers},
  pages={240--243},
  year={2018}
}

@article{wang2020gpgpu,
  title={GPGPU performance estimation with core and memory frequency scaling},
  author={Wang, Qiang and Chu, Xiaowen},
  journal={IEEE Transactions on Parallel and Distributed Systems},
  volume={31},
  number={12},
  pages={2865--2881},
  year={2020},
  publisher={IEEE}
}

@inproceedings{wu2015gpgpu,
  title={GPGPU performance and power estimation using machine learning},
  author={Wu, Gene and Greathouse, Joseph L and Lyashevsky, Alexander and Jayasena, Nuwan and Chiou, Derek},
  booktitle={2015 IEEE 21st international symposium on high performance computer architecture (HPCA)},
  pages={564--576},
  year={2015},
  organization={IEEE}
}

@inproceedings{fan2019predictable,
  title={Predictable GPUs frequency scaling for energy and performance},
  author={Fan, Kaijie and Cosenza, Biagio and Juurlink, Ben},
  booktitle={Proceedings of the 48th International Conference on Parallel Processing},
  pages={1--10},
  year={2019}
}

@article{atomic2013lammps,
  title={Lammps},
  author={Atomic, Large-scale and Simulator, Molecular Massively Parallel},
  journal={available at: http:/lammps. sandia. gov},
  year={2013}
}

@article{van2005gromacs,
  title={GROMACS: fast, flexible, and free},
  author={Van Der Spoel, David and Lindahl, Erik and Hess, Berk and Groenhof, Gerrit and Mark, Alan E and Berendsen, Herman JC},
  journal={Journal of computational chemistry},
  volume={26},
  number={16},
  pages={1701--1718},
  year={2005},
  publisher={Wiley Online Library}
}

@inproceedings{dvfs,
  title={Memory power management via dynamic voltage/frequency scaling},
  author={David, Howard and Fallin, Chris and Gorbatov, Eugene and Hanebutte, Ulf R and Mutlu, Onur},
  booktitle={Proceedings of the 8th ACM international conference on Autonomic computing},
  pages={31--40},
  year={2011}
}

@misc{ecp,
  title ={{ECP proxy apps suite}},
  howpublished={https://proxyapps.exascaleproject.org/ ecp- proxy- apps- suite/.},
  year={2025}
}

@inproceedings{unet,
  title={U-net: Convolutional networks for biomedical image segmentation},
  author={Ronneberger, Olaf and Fischer, Philipp and Brox, Thomas},
  booktitle={Medical image computing and computer-assisted intervention--MICCAI 2015: 18th international conference, Munich, Germany, October 5-9, 2015, proceedings, part III 18},
  pages={234--241},
  year={2015},
  organization={Springer}
}

@inproceedings{keahey2020lessons,
  title={Lessons learned from the chameleon testbed},
  author={Keahey, Kate and Anderson, Jason and Zhen, Zhuo and Riteau, Pierre and Ruth, Paul and Stanzione, Dan and Cevik, Mert and Colleran, Jacob and Gunawi, Haryadi S and Hammock, Cody and others},
  booktitle={2020 USENIX annual technical conference (USENIX ATC 20)},
  pages={219--233},
  year={2020}
}

@inproceedings{farrell2021mlperf,
  title={MLPerf™ HPC: A holistic benchmark suite for scientific machine learning on HPC systems},
  author={Farrell, Steven and Emani, Murali and Balma, Jacob and Drescher, Lukas and Drozd, Aleksandr and Fink, Andreas and Fox, Geoffrey and Kanter, David and Kurth, Thorsten and Mattson, Peter and others},
  booktitle={2021 IEEE/ACM Workshop on Machine Learning in High Performance Computing Environments (MLHPC)},
  pages={33--45},
  year={2021},
  organization={IEEE}
}

@inproceedings{altis-sycl,
  title={Altis-SYCL: Migrating Altis Benchmarking Suite from CUDA to SYCL for GPUs and FPGAs},
  author={Weckert, Christoph and Solis-Vasquez, Leonardo and Oppermann, Julian and Koch, Andreas and Sinnen, Oliver},
  booktitle={Proceedings of the SC'23 Workshops of The International Conference on High Performance Computing, Network, Storage, and Analysis},
  pages={547--555},
  year={2023}
}

@inproceedings{neural_CF,
author = {He, Xiangnan and Liao, Lizi and Zhang, Hanwang and Nie, Liqiang and Hu, Xia and Chua, Tat-Seng},
title = {Neural Collaborative Filtering},
year = {2017},
isbn = {9781450349130},
publisher = {International World Wide Web Conferences Steering Committee},
address = {Republic and Canton of Geneva, CHE},
url = {https://doi.org/10.1145/3038912.3052569},
doi = {10.1145/3038912.3052569},
booktitle = {Proceedings of the 26th International Conference on World Wide Web},
pages = {173–182},
numpages = {10},
location = {Perth, Australia},
series = {WWW '17}
}

@inproceedings{zhang2024improving,
  title={Improving gpu energy efficiency through an application-transparent frequency scaling policy with performance assurance},
  author={Zhang, Yijia and Wang, Qiang and Lin, Zhe and Xu, Pengxiang and Wang, Bingqiang},
  booktitle={Proceedings of the Nineteenth European Conference on Computer Systems},
  pages={769--785},
  year={2024}
}

@incollection{cf,
  title={Collaborative filtering recommender systems},
  author={Schafer, J Ben and Frankowski, Dan and Herlocker, Jon and Sen, Shilad},
  booktitle={The adaptive web: methods and strategies of web personalization},
  pages={291--324},
  year={2007},
  publisher={Springer}
}

@article{bergman2008exascale,
  title={Exascale computing study: Technology challenges in achieving exascale systems},
  author={Bergman, Keren and Borkar, Shekhar and Campbell, Dan and Carlson, William and Dally, William and Denneau, Monty and Franzon, Paul and Harrod, William and Hill, Kerry and Hiller, Jon and others},
  journal={Defense Advanced Research Projects Agency Information Processing Techniques Office (DARPA IPTO), Tech. Rep},
  volume={15},
  pages={181},
  year={2008}
}

@inproceedings{zhao2023power,
  title={Power analysis of nersc production workloads},
  author={Zhao, Zhengji and Rrapaj, Ermal and Bhalachandra, Sridutt and Austin, Brian and Nam, Hai Ah and Wright, Nicholas},
  booktitle={Proceedings of the SC'23 Workshops of the International Conference on High Performance Computing, Network, Storage, and Analysis},
  pages={1279--1287},
  year={2023}
}

@inproceedings{cf_netflex,
  title={Large-scale parallel collaborative filtering for the netflix prize},
  author={Zhou, Yunhong and Wilkinson, Dennis and Schreiber, Robert and Pan, Rong},
  booktitle={Algorithmic Aspects in Information and Management: 4th International Conference, AAIM 2008, Shanghai, China, June 23-25, 2008. Proceedings 4},
  pages={337--348},
  year={2008},
  organization={Springer}
}

@inproceedings{cf_knn,
  title={Design and implementation of movie recommendation system based on knn collaborative filtering algorithm},
  author={Cui, Bei-Bei},
  booktitle={ITM web of conferences},
  volume={12},
  pages={04008},
  year={2017},
  organization={EDP Sciences}
}

@inproceedings{cf_svd,
  title={SVD-based collaborative filtering with privacy},
  author={Polat, Huseyin and Du, Wenliang},
  booktitle={Proceedings of the 2005 ACM symposium on Applied computing},
  pages={791--795},
  year={2005}
}

@inproceedings{cf_autoencoder,
  title={Variational autoencoders for collaborative filtering},
  author={Liang, Dawen and Krishnan, Rahul G and Hoffman, Matthew D and Jebara, Tony},
  booktitle={Proceedings of the 2018 world wide web conference},
  pages={689--698},
  year={2018}
}

@inproceedings{offline-1,
  title={GPU power prediction via ensemble machine learning for DVFS space exploration},
  author={Dutta, Bishwajit and Adhinarayanan, Vignesh and Feng, Wu-chun},
  booktitle={Proceedings of the 15th ACM International Conference on Computing Frontiers},
  pages={240--243},
  year={2018}
}

@inproceedings{offline-2,
  title={Energy efficient real-time task scheduling on CPU-GPU hybrid clusters},
  author={Mei, Xinxin and Chu, Xiaowen and Liu, Hai and Leung, Yiu-Wing and Li, Zongpeng},
  booktitle={IEEE INFOCOM 2017-IEEE Conference on Computer Communications},
  pages={1--9},
  year={2017},
  organization={IEEE}
}

@inproceedings{offline-3,
  title={AccelWattch: A power modeling framework for modern GPUs},
  author={Kandiah, Vijay and Peverelle, Scott and Khairy, Mahmoud and Pan, Junrui and Manjunath, Amogh and Rogers, Timothy G and Aamodt, Tor M and Hardavellas, Nikos},
  booktitle={MICRO-54: 54th Annual IEEE/ACM International Symposium on Microarchitecture},
  pages={738--753},
  year={2021}
}

@inproceedings{offline-4,
  title={Co-run scheduling with power cap on integrated cpu-gpu systems},
  author={Zhu, Qi and Wu, Bo and Shen, Xipeng and Shen, Li and Wang, Zhiying},
  booktitle={2017 IEEE International Parallel and Distributed Processing Symposium (IPDPS)},
  pages={967--977},
  year={2017},
  organization={IEEE}
}

@inproceedings{offline-5,
  title={Power and performance characterization and modeling of GPU-accelerated systems},
  author={Abe, Yuki and Sasaki, Hiroshi and Kato, Shinpei and Inoue, Koji and Edahiro, Masato and Peres, Martin},
  booktitle={2014 IEEE 28th international parallel and distributed processing symposium},
  pages={113--122},
  year={2014},
  organization={IEEE}
}

@inproceedings{offline-6,
  title={GPGPU power modeling for multi-domain voltage-frequency scaling},
  author={Guerreiro, Joao and Ilic, Aleksandar and Roma, Nuno and Tomas, Pedro},
  booktitle={2018 IEEE International Symposium on High Performance Computer Architecture (HPCA)},
  pages={789--800},
  year={2018},
  organization={IEEE}
}

@inproceedings{offline-7,
  title={GPGPU performance and power estimation using machine learning},
  author={Wu, Gene and Greathouse, Joseph L and Lyashevsky, Alexander and Jayasena, Nuwan and Chiou, Derek},
  booktitle={2015 IEEE 21st international symposium on high performance computer architecture (HPCA)},
  pages={564--576},
  year={2015},
  organization={IEEE}
}

@article{offline-8,
  title={Modeling and decoupling the GPU power consumption for cross-domain DVFS},
  author={Guerreiro, Jo{\~a}o and Ilic, Aleksandar and Roma, Nuno and Tom{\'a}s, Pedro},
  journal={IEEE Transactions on Parallel and Distributed Systems},
  volume={30},
  number={11},
  pages={2494--2506},
  year={2019},
  publisher={IEEE}
}

@article{offline-9,
  title={A survey and measurement study of GPU DVFS on energy conservation},
  author={Mei, Xinxin and Wang, Qiang and Chu, Xiaowen},
  journal={Digital Communications and Networks},
  volume={3},
  number={2},
  pages={89--100},
  year={2017},
  publisher={Elsevier}
}

@article{offline-10,
  title={GPGPU performance estimation with core and memory frequency scaling},
  author={Wang, Qiang and Chu, Xiaowen},
  journal={IEEE Transactions on Parallel and Distributed Systems},
  volume={31},
  number={12},
  pages={2865--2881},
  year={2020},
  publisher={IEEE}
}

@inproceedings{offline-11,
  title={Gpgpu performance estimation for frequency scaling using cross-benchmarking},
  author={Wang, Qiang and Liu, Chengjian and Chu, Xiaowen},
  booktitle={Proceedings of the 13th Annual Workshop on General Purpose Processing Using Graphics Processing Unit},
  pages={31--40},
  year={2020}
}

@inproceedings{hybrid-1,
  title={Optimal GPU Frequency Selection using Multi-Objective Approaches for HPC Systems},
  author={Ali, Ghazanfar and Bhalachandra, Sridutt and Wright, Nicholas J and Side, Mert and Chen, Yong},
  booktitle={2022 IEEE High Performance Extreme Computing Conference (HPEC)},
  pages={1--7},
  year={2022},
  organization={IEEE}
}

@inproceedings{hybrid-2,
  title={Performance-Aware Energy-Efficient GPU Frequency Selection using DNN-based Models},
  author={Ali, Ghazanfar and Side, Mert and Bhalachandra, Sridutt and Wright, Nicholas J and Chen, Yong},
  booktitle={Proceedings of the 52nd International Conference on Parallel Processing},
  pages={433--442},
  year={2023},
  publisher={ACM}
}

@inproceedings{hybrid-3,
  title={A Data-Driven Frequency Scaling Approach for Deadline-aware Energy Efficient Scheduling on Graphics Processing Units (GPUs)},
  author={Ilager, Shashikant and Muralidhar, Rajeev and Ramamohanarao, Kotagiri and Buyya, Rajkumar},
  booktitle={2020 20th IEEE/ACM International Symposium on Cluster, Cloud and Internet Computing (CCGRID)},
  pages={579--588},
  year={2020},
  organization={IEEE}
}

@inproceedings{hybrid-4,
  title={GreenGPU: A Holistic Approach to Energy Efficiency in GPU-CPU Heterogeneous Architectures},
  author={Ma, Kai and Li, Xue and Chen, Wei and Zhang, Chi and Wang, Xiaorui},
  booktitle={2012 41st International Conference on Parallel Processing},
  pages={48--57},
  year={2012},
  organization={IEEE}
}

@inproceedings{hybrid-5,
  title={Power capping of CPU-GPU heterogeneous systems through coordinating DVFS and task mapping},
  author={Komoda, Toshiya and Hayashi, Shingo and Nakada, Takashi and Miwa, Shinobu and Nakamura, Hiroshi},
  booktitle={2013 IEEE 31st International Conference on Computer Design (ICCD)},
  pages={349--356},
  year={2013},
  organization={IEEE}
}

@inproceedings{hybrid-6,
  title={Coordinated energy management in heterogeneous processors},
  author={Paul, Indrani and Ravi, Vignesh and Manne, Srilatha and Arora, Manish and Yalamanchili, Sudhakar},
  booktitle={SC '13: Proceedings of the International Conference on High Performance Computing, Networking, Storage and Analysis},
  pages={1--12},
  year={2013},
  organization={IEEE}
}

@inproceedings{hybrid-7,
  title={Harmonia: Balancing Compute and Memory Power in High-Performance GPUs},
  author={Paul, Indrani and Huang, Wei and Arora, Manish and Yalamanchili, Sudhakar},
  booktitle={Proceedings of the 42nd Annual International Symposium on Computer Architecture},
  pages={54--65},
  year={2015},
  publisher={ACM}
}

@inproceedings{hybrid-8,
  title={Multi-kernel Auto-Tuning on GPUs: Performance and Energy-Aware Optimization},
  author={Guerreiro, João and Ilic, Aleksandar and Roma, Nuno and Tomás, Pedro},
  booktitle={2015 23rd Euromicro International Conference on Parallel, Distributed, and Network-Based Processing},
  pages={438--445},
  year={2015},
  organization={IEEE}
}

@inproceedings{hybrid-9,
  title={Dynamic GPGPU Power Management Using Adaptive Model Predictive Control},
  author={Majumdar, Abhinandan and Piga, Leonardo and Paul, Indrani and Greathouse, Joseph L and Huang, Wei and Albonesi, David H},
  booktitle={2017 IEEE International Symposium on High Performance Computer Architecture (HPCA)},
  pages={613--624},
  year={2017},
  organization={IEEE}
}

@inproceedings{hybrid-10,
  title={Predictable GPUs Frequency Scaling for Energy and Performance},
  author={Fan, Kaijie and Cosenza, Biagio and Juurlink, Ben},
  booktitle={Proceedings of the 48th International Conference on Parallel Processing},
  pages={1--10},
  year={2019},
  publisher={ACM}
}

@article{hybrid-11,
  title={DVFS-aware application classification to improve GPGPUs energy efficiency},
  author={Guerreiro, João and Ilic, Aleksandar and Roma, Nuno and Tomás, Pedro},
  journal={Parallel Computing},
  volume={83},
  pages={93--117},
  year={2019},
  publisher={Elsevier}
}

@article{hybrid-12,
  title={Coordinated batching and DVFS for DNN inference on GPU accelerators},
  author={Nabavinejad, Seyed Morteza and Reda, Sherief and Ebrahimi, Masoumeh},
  journal={IEEE transactions on parallel and distributed systems},
  volume={33},
  number={10},
  pages={2496--2508},
  year={2022},
  publisher={IEEE}
}

@inproceedings{sim-1,
  title={Equalizer: Dynamic tuning of gpu resources for efficient execution},
  author={Sethia, Ankit and Mahlke, Scott},
  booktitle={2014 47th Annual IEEE/ACM International Symposium on Microarchitecture},
  pages={647--658},
  year={2014},
  organization={IEEE}
}

@inproceedings{sim-2,
  title={The CRISP performance model for dynamic voltage and frequency scaling in a GPGPU},
  author={Nath, Rajib and Tullsen, Dean},
  booktitle={Proceedings of the 48th international symposium on microarchitecture},
  pages={281--293},
  year={2015}
}

@inproceedings{sim-3,
  title={Grape: Minimizing energy for gpu applications with performance requirements},
  author={Santriaji, Muhammad Husni and Hoffmann, Henry},
  booktitle={2016 49th Annual IEEE/ACM International Symposium on Microarchitecture (MICRO)},
  pages={1--13},
  year={2016},
  organization={IEEE}
}

@inproceedings{online-1,
  title={Indicator-directed dynamic power management for iterative workloads on GPU-accelerated systems},
  author={Zou, Pengfei and Li, Ang and Barker, Kevin and Ge, Rong},
  booktitle={2020 20th IEEE/ACM International Symposium on Cluster, Cloud and Internet Computing (CCGRID)},
  pages={559--568},
  year={2020},
  organization={IEEE}
}

@inproceedings{online-2,
  title={Improving gpu energy efficiency through an application-transparent frequency scaling policy with performance assurance},
  author={Zhang, Yijia and Wang, Qiang and Lin, Zhe and Xu, Pengxiang and Wang, Bingqiang},
  booktitle={Proceedings of the Nineteenth European Conference on Computer Systems},
  pages={769--785},
  year={2024}
}

@inproceedings{model-1,
  title={AccelWattch: A power modeling framework for modern GPUs},
  author={Kandiah, Vijay and Peverelle, Scott and Khairy, Mahmoud and Pan, Junrui and Manjunath, Amogh and Rogers, Timothy G and Aamodt, Tor M and Hardavellas, Nikos},
  booktitle={MICRO-54: 54th Annual IEEE/ACM International Symposium on Microarchitecture},
  pages={738--753},
  year={2021}
}

@article{model-2,
  title={GPGPU performance estimation with core and memory frequency scaling},
  author={Wang, Qiang and Chu, Xiaowen},
  journal={IEEE Transactions on Parallel and Distributed Systems},
  volume={31},
  number={12},
  pages={2865--2881},
  year={2020},
  publisher={IEEE}
}

@inproceedings{model-3,
  title={GPGPU power modeling for multi-domain voltage-frequency scaling},
  author={Guerreiro, Joao and Ilic, Aleksandar and Roma, Nuno and Tomas, Pedro},
  booktitle={2018 IEEE International Symposium on High Performance Computer Architecture (HPCA)},
  pages={789--800},
  year={2018},
  organization={IEEE}
}

@article{model-4,
  title={GPGPU power estimation with core and memory frequency scaling},
  author={Wang, Qiang and Chu, Xiaowen},
  journal={ACM SIGMETRICS Performance Evaluation Review},
  volume={45},
  number={2},
  pages={73--78},
  year={2017},
  publisher={ACM New York, NY, USA}
}

@inproceedings{model-5,
  title={Improving gpu energy efficiency through an application-transparent frequency scaling policy with performance assurance},
  author={Zhang, Yijia and Wang, Qiang and Lin, Zhe and Xu, Pengxiang and Wang, Bingqiang},
  booktitle={Proceedings of the Nineteenth European Conference on Computer Systems},
  pages={769--785},
  year={2024}
}

@inproceedings{model-6,
  title={Power and performance characterization and modeling of GPU-accelerated systems},
  author={Abe, Yuki and Sasaki, Hiroshi and Kato, Shinpei and Inoue, Koji and Edahiro, Masato and Peres, Martin},
  booktitle={2014 IEEE 28th international parallel and distributed processing symposium},
  pages={113--122},
  year={2014},
  organization={IEEE}
}

@inproceedings{nn_better_1,
  title={GPU power prediction via ensemble machine learning for DVFS space exploration},
  author={Dutta, Bishwajit and Adhinarayanan, Vignesh and Feng, Wu-chun},
  booktitle={Proceedings of the 15th ACM International Conference on Computing Frontiers},
  pages={240--243},
  year={2018}
}

@inproceedings{nn_better_2,
  title={GPGPU performance and power estimation using machine learning},
  author={Wu, Gene and Greathouse, Joseph L and Lyashevsky, Alexander and Jayasena, Nuwan and Chiou, Derek},
  booktitle={2015 IEEE 21st international symposium on high performance computer architecture (HPCA)},
  pages={564--576},
  year={2015},
  organization={IEEE}
}

@inproceedings{cpu-gpu,
  title={Coordinated energy management in heterogeneous processors},
  author={Paul, Indrani and Ravi, Vignesh and Manne, Srilatha and Arora, Manish and Yalamanchili, Sudhakar},
  booktitle={Proceedings of the International Conference on High Performance Computing, Networking, Storage and Analysis},
  pages={1--12},
  year={2013}
}

@inproceedings{che2009rodinia,
  title={Rodinia: A benchmark suite for heterogeneous computing},
  author={Che, Shuai and Boyer, Michael and Meng, Jiayuan and Tarjan, David and Sheaffer, Jeremy W and Lee, Sang-Ha and Skadron, Kevin},
  booktitle={2009 IEEE international symposium on workload characterization (IISWC)},
  pages={44--54},
  year={2009},
  organization={Ieee}
}

@inproceedings{hong2010integrated,
  title={An integrated GPU power and performance model},
  author={Hong, Sunpyo and Kim, Hyesoon},
  booktitle={Proceedings of the 37th annual international symposium on Computer architecture},
  pages={280--289},
  year={2010}
}

@inproceedings{lee2011improving,
  title={Improving throughput of power-constrained GPUs using dynamic voltage/frequency and core scaling},
  author={Lee, Jungseob and Sathisha, Vijay and Schulte, Michael and Compton, Katherine and Kim, Nam Sung},
  booktitle={2011 International Conference on Parallel Architectures and Compilation Techniques},
  pages={111--120},
  year={2011},
  organization={IEEE}
}

@techreport{strawman,
  title={A strawman for an hpc powerstack},
  author={Cantalupo, Christopher and Eastep, Jonathan and Jana, Siddhartha and Kondo, Masaaki and Maiterth, Matthias and Marathe, Aniruddha and Patki, Tapasya and Rountree, Barry and Sakamoto, Ryuichi and Schulz, Martin and others},
  year={2018},
  institution={Intel Corporation (United States); Lawrence Livermore National Lab.(LLNL~…}
}


\end{document}